\documentclass[10pt]{article}

\usepackage{amssymb}          
\usepackage{graphicx}         
\usepackage[latin1]{inputenc} 
\usepackage{amsmath}          

\usepackage{amsfonts}

\usepackage{latexsym}
\usepackage{graphicx}
\usepackage{color}

\newtheorem{proposition}{Proposition}
\newtheorem{definition}{Definition}
\newtheorem{theorem}{Theorem}
\newtheorem{corollary}{Corollary}
\newtheorem{lemma}{Lemma}
\newtheorem{prob}{Problem}
\newcommand{\proof}{\noindent {\bf Proof. }}
\newcommand{\qed}{\hfill $\Box$ \vskip 2ex}

\newcommand{\vv}[1]{``#1''}

\newcommand{\ket}[1]{|#1\rangle}
\newcommand{\bra}[1]{\langle#1|}

\newcommand{\Hi}{\mathcal{H}}

\newcommand{\C}{\mathbb{C}}

\newcommand{\R}{\mathbb{R}}
\newcommand{\cvd}{\hfill $\Box$ \vskip 2ex}

\newcommand{\tr}{\textrm{tr}}

\newcommand{\Herm}{\textrm{Herm}}

\newcommand{\beq}{\begin{equation}}
\newcommand{\eeq}{\end{equation}}
\newcommand{\beqa}{\begin{eqnarray}}
\newcommand{\eeqa}{\end{eqnarray}}
\newcommand{\beqan}{\begin{eqnarray*}}
\newcommand{\eeqan}{\end{eqnarray*}}

\title{Almost Global Stochastic Feedback Stabilization of Conditional Quantum Dynamics \thanks{Partially supported by the ministry of higher education of Italy (MIUR), under projects {\em Control, Optimization and Stability of Nonlinear Systems: Geometric and Analitic Methods} and {\em Identification and Control of Industrial Systems}.}}
\author{Claudio Altafini\thanks{SISSA, via
Beirut 4, 34014 Trieste, Italy ({\tt altafini@sissa.it}).} \and
Francesco Ticozzi\thanks{Dipartimento di Ingegneria
dell'Informazione, Universit\`a di Padova, via Gradenigo 6/B,
35131 Padova, Italy ({\tt ticozzi@dei.unipd.it}).}}
\date{\today}

\begin{document}

\maketitle
\begin{abstract}

We propose several parametrization-free solutions to the problem
of quantum state reduction control by means of continuous
measurement and smooth quantum feedback.
In particular, we design a feedback law for which almost global stochastic feedback stabilization can be proved analytically by means of Lyapunov techinques.
This synthesis arises
very naturally from the physics of the problem, as it relies on the variance associated with the quantum filtering
process.

\end{abstract}


\section{Introduction}

Experimental techniques in quantum optics permit nowadays to
continuously monitor and modify the dynamics of a cloud of cold 
atoms confined in an optical cavity \cite{mabuchi-science}. 
The knowledge of the system state can be described by a 
conditional dynamical equation, the so-called Stochastic 
Master Equation (SME), obtained from a suitable quantum 
system-field interaction model by means of non-commutatitive filtering theory 
\cite{wiseman-milburn,belavkin-filtering}, 
and based on monitoring the
outgoing field from the cavity. 
The stochastic 
perturbation induced by the indirect measurement process produces
an effective {\em dynamical reduction model} \cite{Adler1}. In
other words, it makes the system state converge to one of the
maximal information, pure states for the system physical
observable interacting with the field.

If we have a second controllable field, acting as a time dependent
Hamiltonian perturbation, then we can use the real-time estimate
of the system state to modify the state reduction process. One
possible use of such a feedback control strategy can be {\em
choosing to which pure state of the monitoring observable the
system will converge.} Beside possible interest for quantum
measurement theory, the overall strategy can then be seen as a
technique for engineering quantum {\em state preparation}.

From a control theoretic viewpoint, the problem is doubtless
challenging.
The SME is a nonlinear affine in control matrix Stochastic Differential
Equation (SDE) living in the convex cone of positive semidefinite
$N$-dimensional Hermitian matrices.
In the particular case of perfect measurement efficiency and
maximal information on the initial condition, a SME turns out to
be equivalent to a Stochastic Schr\"odinger Equation (SSE), i.e. a
vector-valued, norm-preserving diffusion in $\C^N$ \cite{Adler1,bouten-sse}.
Influencing the open loop state reduction by means of the measurement is therefore a stochastic nonlinear feedback stabilization problem.
Partial solutions to this problem were presented for example in \cite{wang1} for 2 level SSE and, in more general terms, in \cite{vanhandel} based on a (convex) numerical Lyapunov design.
This last solution, however, suffers from scalability problems as the dimension of the system grows, since it is based on explicit parametrizations which grow with $N^2 $ if $ N$ is the dimension of the system.

For nonlinear (vector) SDE, most of the results on state feedback stabilization are due to Florchinger \cite{Florchinger1,Florchinger2,Florchinger3} (see also e.g. \cite{Deng1,Deng2,Battilotti1} for other possible approaches) and consists of extensions of Lyapunov-based techniques, like the Jurdjevic-Quinn condition, control Lyapunov function constructions, passivity-based methods and so on, to the stochastic case.

For our SME, these systematic construction methods have a limited success, and allow only to attain local stabilization in the particular case of SSE.
The feedback that achieves it is the simple linear feedback already used for deterministic unitary feedback stabilization of quantum ensembles \cite{Cla-qu-ens-feeb1}.
However, owing to the nature of the problem, local convergence results are of no practical interest.
It is the structure of the problem itself that suggests how to improve the design: the state reduction encoded in the SME is ``certified'' by the variance of the continuous measurement process, a multiequilibrium Lyapunov-like function (with $N$ equilibria corresponding to the $N$ eigenvectors of the observable being measured) which is also a Morse function and can be used to enlarge the region of attraction of the controller.
We shall in particular discuss two nonlinear feedback laws induced by the use of the variance, both more effective than the linear controller.
Both allow for simple explicit proofs of convergence: one corresponds to a closed loop stochastic generator which is a sum of squares, the other to the square of a sum.
The former achieves almost global stabilization for the perfect efficiency case, but cannot cancel all invariant sets of the dynamics in the more general SME and thus suffers from similar problems as the linear controller.
The latter instead corresponds to a feedback stabilization design for the SME which is almost global, up to the $N-1 $ isolated repulsive critical points (the remaining $ N-1$ eigenvectors of the measured observable).
This feedback strategy extends the idea of ``symmetry breaking'' enunciated in \cite{vanhandel}. Indeed it works by using the uncontrollable part of the drift term to evade from the zero-control locus.
We also show how the rate of convergence can be modified by tuning opportunely a pair of gains corresponding to the relative weights given to the controlled and uncontrolled parts of the stochastic generator.

\section{Model formulation and background material }
We need to recall some basics of quantum mechanics, quantum
filtering and stochastic stability theory we will use later on.
For an excellent introductory exposition to the statistical description
of quantum systems see e.g. \cite{maassen-qp}. More
details can be found in e.g. \cite{meyer,parthasarathy} and
references therein.
For the theory of stochastic stability, main references are \cite{Arnold1,Khasminskiy1}, while for stochastic feedback stabilization we shall make use of the works of Florchinger \cite{Florchinger1,Florchinger2,Florchinger3}.
Needless to say, the paper \cite{vanhandel} presents similar control-theoretic perspective on most of the material mentioned below.

\subsection{Quantum finite dimensional systems}


In the standard statistical formulation of quantum mechanics
\cite{holevo,sakurai}, to each quantum system is associated an
Hilbert space $\Hi$, whose dimension depends essentially on the
observable quantities we want to describe. In fact, physical
observables are modelled as self-adjoint operators in the Hilbert
space, the set of possible outcomes they can assume being their
spectrum. In what follows, we will consider only {\em observables}
with finite spectrum, thus represented as Hermitian matrices $C
\in\Herm$ acting on $\C^N$.

Our knowledge of the system will be represented by a {\em density
matrix} $\rho$ belonging to the convex set
\[
\label{densityset}
\mathcal{M}= \{ \rho = \rho^\dagger \geq 0\text{ s. t. } \tr( \rho) = 1 \}.
\]
The extremals of $\mathcal{M}$ are the one dimensional orthogonal
projections. These are called \emph{pure states}, and are
equivalent to unit vectors in $\mathcal{H}_S$ up to an overall
phase factor, by setting
$\rho=\ket{\psi}\bra{\psi}=\langle\psi,\cdot\rangle \psi$. We
will use Dirac's notation for vectors: $\ket{\psi}\in\Hi$,
$\bra{\psi}\in\Hi^\dag$. Unit vectors in $\Hi$ will be thus called
{\em state vectors}. The usual notations $\langle\psi,\phi\rangle$
and $\langle\psi,C\phi\rangle$ will be replaced by
$\langle{\psi}|{\phi}\rangle$, $\bra{\psi}C\ket{\phi}$.

Consider an observable $C$ and its spectral decomposition
$C=\sum_ic_iP_i$, where $\{P_i\}$ is a {\em spectral family} of
orthogonal projectors summing to the identity. Assume one can
perform an ideally instantaneous measurement of $C$. The
probability to obtain $c_i$ as an outcome is then given by
\[
\label{probability}
p(c_i)=\tr(P_i\rho).
\]
Thus, a density
matrix determines also the expectation value of an observable
\[
\label{expectationvalue}
\langle C \rangle=\tr(\rho C).
\]
If $c_i$ is the observed outcome, the conditioned density matrix is
given by the L\"uders-von Neumann postulate as
\beq
\label{luders}
\rho_i=\frac{P_i\rho P_i}{\tr(P_i\rho)}.
\eeq

We will assume throughout the paper to work in measurement units
such that $\hbar=1$, where $\hbar$ is the Plank constant divided by
$2\pi$. In absence of measurements, the time evolution of an
isolated quantum system is driven by the Hamiltonian $H$, i.e., the
energy observable, as specified by:
\[
\label{eq:liouvillian} \frac{d}{dt}\rho=-i[H,\rho].
\]
Notice that the
evolution of the unobserved system is deterministic, and, if
$\rho=\ket{\psi}\bra{\psi}$ is pure, it is equivalent to the
Scr\"odinger equation
\[
\label{eq:schrodinger}
\frac{d}{dt}\ket{\psi}=-iH\ket{\psi}.
\]
Beside of these basic postulates and definitions, to tackle our
main problem we will need more sophisticated tools to deal with
continuous-time measurement and subsequent state conditioning.

\subsection{Continuous Measurement and Filtering Equation}
\label{sec:model-SME-pres}

For explicit derivations and more detailed discussions of the
following topics, we refer to e.g.
\cite{belavkin-filtering, belavkin-nondemolition, bouten-sse,
vanhandel}.

In the description of a classical uncertain system, observable
quantities are represented by real \emph{random variables} defined
on a suitable probability space $(\Omega,\Sigma,\mathbb{P})$. The
state of the system, i.e., our knowledge about it, is subsumed in
the probability measure $\mathbb P$. The observables form a
commutative algebra, $L^\infty(\Omega,\Sigma,\mathbb P;\R)$.

The quantum setting presented in the previous section can be
interpreted as a non commutative generalization of a classical,
discrete probability space \cite{maassen-qp}. The need for
non-commutativity emerges experimentally, motivating the standard
axiomatic formulation of the theory and being essentially captured
by the canonical commutation relations \cite{sakurai}. Briefly,
quantum \emph{observables}, or non commutative random variables,
form a (generally non commutative) Von Neumann algebra
$\mathcal{A}$ and \emph{events} are represented by orthogonal
projections $\mathcal{E}\subset\mathcal{A}$ in the algebra. A
\emph{generalized probability measure} on $\mathcal{E}$ is needed
to compute probabilities of events.

The finite dimensional setting we are working in leads to a
concrete representation of the above abstract notions. We can
identify $\mathcal{A}$ with $\Herm$, and the set of generalized
probability densities with $\mathcal{M}$, determining probabilities through
$p(E)=\tr(\rho E),\,E\in\mathcal{E}$.

One can then apply \emph{quantum filtering theory} to obtain a
conditional equation on $ {\cal A} $ for the dynamics
\cite{belavkin-filtering}. It essentially plays the role of the
classical Kushner-Stratonovich equation.  Thus assume we are
continuously monitoring the observable $C$ for the system of
interest. In quantum optics, this can be accomplished e.g. for an
angular momentum observable $C$ by an homodyne detection
experimental setup \cite{wiseman-milburn}\footnote{In general,
with homodyne detection one makes continuous-time measurements of
generalized operators of the form:
\[\mathcal{R}[\rho]=L\rho+\rho L^\dag,\]
measuring the outgoing field from an optical cavity where we
confine the system. The operator $L$ depends on the system-field
interaction occurring in the cavity. We will specialize to the
case of Hermitian operators $L=L^\dag=C$.}. 
Since the observed $C$ is
time-invariant, we are conditioning the dynamics on the
observation of a {\em commuting} quantum stochastic process, that
leads to a dynamical equation driven by classical white noise (see
e.g. \cite{belavkin-nondemolition}).

Let $ ( \Omega, {\cal E}, P ) $ a (classical) probability space
and $ \{ W_t , t\in \mathbb{R}^+ \} $ a standard $
\mathbb{R}$-valued Wiener process defined on this space. The
homodyne detection measurement record can be written as the output
of a stochastic dynamical system of the form: \beq
\label{photocurrent} dY_t=\sqrt{\eta}\tr(\rho_t C)dt+dW_t, \eeq
where $0\leq\eta\leq 1$ represents the efficiency of the
measurement. Denote with $\mathcal{E}_t$ the filtration associated
to $ \{ W_t , t\in \mathbb{R}^+ \} $.

Then one can derive the filtering equation determining the
conditional evolution of the state for the measurement record
(\ref{photocurrent}), the Quantum Filtering or Stochastic Master
Equation (SME) \`a la It\^o:
\begin{equation}
\begin{split}
d\rho_t & = ({\cal F}( H,\rho_t) + {\cal D}(C, \rho_t)) dt  + {\cal G} (C, \rho_t) d W_t \\
& = \left(-i[H,\rho_t]+\mu C\rho_t C -\frac{\mu}{2}(C^2\rho_t+\rho_t
C^2)\right)dt+\sqrt{\mu \eta }(C\rho_t+\rho_t C-2 \tr( C \rho_t) \rho_t)dW_t,
\end{split}
\label{eq:SME-Ito1}
\end{equation}
where $ {\cal F} $ represent the Hamiltonian part, with $ H$ given
by a drift and a (bilinear) control part $ H = H_a + u H_b $, $
{\cal D} $ and $ {\cal G} $ are the drift and diffusion parts of the
weak measurement performed along the observable $ C = C ^\dagger$.
The parameter $ \mu > 0 $ represents the strength of the
measurement.

Here $ \rho_t $, the $ \mathcal{M}$-valued solution of
\eqref{eq:SME-Ito1} given a constant initial condition $\rho_0$,
that can be written explicitly as
\[
\begin{split}
 \rho_t & = \Phi ( \rho_0, t, 0 ) , \quad \rho_0 \in \mathcal{M} \\
& = \rho_0 + \int_0^t  \left( {\cal F}( H,\rho_s) + {\cal D}(C, \rho_s) \right) ds +
\int_0^t  {\cal G} (C, \rho_s) d W_s,
\end{split}
\]
exists, is unique, adapted to the filtration $\mathcal{E}_t$ and $
\mathcal{M}$-invariant by construction, see
\cite{belavkin-filtering,vanhandel}.

Considering \eqref{photocurrent} and \eqref{eq:SME-Ito1} together,
one can recognize the basic structure of a {\em Kalman-Bucy}
filter. Since $\langle C \rangle_t=\tr(C \rho_t)$ is the
expectation of $Y_t$ at
time $t$, in \eqref{photocurrent} $dW_t$ plays the role of {\em
innovation process} in a filtering model. Other correspondences
with the classical setting have been discussed and highlighted in
e.g. \cite{mabuchi-quantumclassical}.

We denote by $ {\cal L} $ the infinitesimal generator \`a la It\^o
associated with the SME \eqref{eq:SME-Ito1}, written in a
``symmetrized'' fashion \beq
\begin{split}
{\cal L} \cdot & =
\frac{1}{2} \left(  ({\cal F}( H,\rho_t)  + {\cal D} ( C, \rho_t )) \frac{\partial \, \cdot }{\partial \rho} + \frac{\partial \, \cdot }{\partial \rho} ({\cal F}( H,\rho_t)  + {\cal D} ( C, \rho_t )) \right. \\
& \left. + {\cal G} ^2(C, \rho_t) ) \frac{\partial^2 \, \cdot
}{\partial \rho^2} + \frac{\partial^2 \, \cdot }{\partial \rho^2}
{\cal G}^2 (C, \rho_t) ) \right).
\end{split}
\label{eq:inf-gen-Ito} \eeq

Consider now the case of perfect detection efficiency $ \eta =1 $.
In this case, a pure $ \rho_0 $ remains pure throughout the
evolution. In fact, recalling that $\rho_t=\ket{\psi_t}\bra{\psi_t}$ if
and only if $\tr(\rho_t^2)=1$, it suffices to prove the following.

\begin{lemma} Consider \eqref{eq:SME-Ito1} with $\rho_t$ a pure state and $\eta=1$. Then $ d\textrm{{\em tr}}(\rho_t^2)=0. $
\end{lemma}\proof Using Ito's rule, we have:
\[
\begin{split}
d \tr (\rho_t^2 ) & = \tr ( 2 \rho_t d \rho_t + ( d \rho_t)^2 ) \\
& = \tr \left( 2 \rho_t \left(  {\cal F} + {\cal D} \right) d t + 2
\rho_t {\cal G} d W \right) + \tr \left( {\cal G} ^2 dt \right)\\
& = \mu \tr \left( 2 \rho_t \left( C \rho_t C - \frac{1}{2} C^2 \rho_t - \frac{1}{2} \rho_t C^2 \right) \right) dt
+ \mu \eta \tr \left( \left( (C\rho_t + \rho_t C  - 2 \langle C \rangle_t)   \rho_t \right) ^2 \right) dt \\
& + 4 \sqrt{\mu \eta }  \tr \left(  (C - \langle C \rangle_t)   \rho_t
^2 \right) d W_t \\
& = \mu \tr \left( 2 ( 1+ \eta ) \rho_t C \rho_t C - 2 ( 1- \eta ) C^2 \rho_t^2
+  \eta \langle C \rangle_t ^2 \rho_t^2 - 8 \eta  \langle C \rangle_t  \rho_t^2 \right) dt \\
& +  4 \sqrt{\mu \eta }  \tr \left(  (C - \langle C \rangle_t)   \rho_t
^2 \right) d W_t .
\end{split}
\]
If $ \eta =1 $ the term $ C^2 \rho_t^2 $ disappears and
\beq
\begin{split}
\tr \left( 2 \rho_t (  {\cal F} + {\cal D}) + {\cal G} ^2 \right)  & = \mu \tr (4 C \rho_t C \rho_t + 4 \langle C \rangle_t^2  \rho_t ^2 - 8  \langle C \rangle_t \rho_t^2 )  \\
& = 4 \mu \tr \left( \left( ( C  - \langle C \rangle_t)  \rho_t \right)^2  \right) .
\end{split}
\label{eq:tr-rho^2-1}
\eeq
The assumption of starting with a pure state $\rho_t=\ket{\psi_t}\bra{\psi_t}$
implies for example $\rho_t^2=\rho_t$, $\tr(\rho_t^2)=1$ and $\tr(\rho_t
C\rho_t C)=\bra{\psi_t}C\ket{\psi_t}\bra{\psi_t}C\ket{\psi_t}=\langle C
\rangle_t^2$. Hence
\beq
\begin{split}
4 \mu \tr \left( \left( ( C  - \langle C \rangle_t)  \rho_t \right)^2
\right)&= 4\mu \tr(C\rho_t C\rho_t-2\langle C \rangle_t C\rho_t^2+\langle
C
\rangle_t^2\rho_t^2)\\
&=4\mu \left(\langle C \rangle_t^2-2\langle C
\rangle_t\tr(C\rho_t)+\langle C \rangle_t^2\right)\\
&=0
\end{split}
\eeq and, likewise, \beq \tr \left( ( C  - \langle C \rangle_t)
\rho_t^2 \right) =0 \label{eq:eta=1=fluct=0} \eeq \cvd

Thus, the SME \eqref{eq:SME-Ito1} becomes equivalent to a
Stochastic Schr\"odinger Equation (SSE) of the form
\cite{bouten-sse}:
\beq
\label{eq:SSE}
d\ket{\psi_t}=\left(-iH-\frac{\mu}{2}(C-\langle C
\rangle_t)^2\right)\ket{\psi_t}dt+\sqrt{\mu}(C-\langle C
\rangle_t)\ket{\psi_t}dW_t .
\eeq
In particular (see \cite{Cla-qu-ens-feeb1} for details), the state space in this
case, call it $ {\cal S} $, reduces to a homogeneous space of the Lie group $ U(N)$:
\[
{\cal S} = U(N) / ( U(N-1) \times U(1) ) \subset \mathcal{M},
\]
of $ {\rm dim} ({\cal S}) = N^2 - N $.

Equations of the form \eqref{eq:SSE} have been proposed  as
extensions to standard quantum mechanics in order to give a
dynamical model for the after measurement \vv{state collapse},
i.e. postulate \eqref{luders} (see e.g. \cite{Adler1} and
references therein).


\subsection{Elements of stochastic stability}
Consider $\rho_d$ an equilibrium solution of \eqref{eq:SME-Ito1},
i.e. $ \rho_d \in \mathcal{M} $: $ {\cal F}( H , \rho_d ) + {\cal D} ( C,
\rho_d ) = {\cal G} (C, \rho_d ) =0 $.

\begin{definition}
The equilibrium $ \rho_d $ of the SME \eqref{eq:SME-Ito1} is said to be
\begin{enumerate}
\item {\em stable in probability} if for any $ s\geqslant 0 $ and $ \epsilon \geqslant 0 $
\beq
\lim_{\rho_0 \to \rho_d } P \left( \sup \left| \Phi ( \rho_0, t, s )- \rho_d  \right| > \epsilon \right) =0 ;
\label{eq:def-stable}
\eeq
\item {\em locally asymptotically stable in probability} if \eqref{eq:def-stable} holds and
\beq
\lim_{\rho_0 \to \rho_d } P \left( \lim_{t\to \infty}  \left| \Phi ( \rho_0, t, s )- \rho_d  \right| =0 \right) =1 ;
\label{eq:def-loc-as-stable}
\eeq
\item {\em almost globally asymptotically stable in probability} if \eqref{eq:def-stable} holds and \eqref{eq:def-loc-as-stable} is true $ \forall \; \rho_0 \in \mathcal{M} $ except for at most a finite number of isolated points of $ \mathcal{M} $;
\item {\em globally asymptotically stable in probability} if \eqref{eq:def-stable} holds and
\beq
 P \left( \lim_{t\to \infty}  \left| \Phi ( \rho_0, t, s )- \rho_d  \right| =0 \right) =1 .
\label{eq:def-glob-as-stable}
\eeq
\end{enumerate}
\end{definition}
We shall make use of the following Lyapunov conditions.

\begin{theorem}
\label{thm-Lyap-stab1}
Denote by $\mathcal{B}_\mathcal{M} $ the intersection of an open neighborhood $ \mathcal{B} \in {\rm Herm} $ with the set of density operators: $ \mathcal{B}_\mathcal{M} = \mathcal{B} \cap \mathcal{M}$.
Assume $ \exists $ a $ \mathbb{R}$-valued $ V \in C^2(\mathcal{B}_\mathcal{M} , \, \mathbb{R}) $ with $ V( \rho_d ) =0 $, $ V( \mathcal{B}_\mathcal{M} \smallsetminus \{ \rho_d \} )>0 $ and such that $ {\cal L} V_t = {\cal L} V (\rho_t ) \leqslant 0 $ ( resp.  $ {\cal L} V_t  < 0 $) $ \forall \; \rho_t \in \mathcal{B}_\mathcal{M} \smallsetminus \{ \rho_d \} $.
Then $ \rho_d $ is locally stable (resp. locally asymptotically stable) in probability.
\end{theorem}

Since \eqref{eq:SME-Ito1} is invariant in $\mathcal{M}$, the restriction of a full neighborhood to $ \mathcal{B}_\mathcal{M} $ is not altering the standard proof of this result (reported for example in \cite{Khasminskiy1}).

Just like in the deterministic case, a well-established version of the LaSalle's invariance principle provides the $ \omega$-limit set of a stable stochastic process.
\begin{theorem}
\label{thm:LaSalle}
(\cite{Kushner1})
If $ \exists $ a Lyapunov function $ V \in C^2 ( \mathcal{M}, \, \mathbb{R}) $ such that $ {\cal L} V _t  \leqslant 0 $ $ \forall \; \rho_t \in \mathcal{M}$, then the solution $ \rho_t $ of \eqref{eq:SME-Ito1} tends with probability 1 to the largest invariant set whose support is contained in $ {\cal N} = \{ \rho_t \in \mathcal{M} \text{ s. t. } {\cal L}V_t  = 0 \; \;  \forall \; t\geqslant 0 \} $.
\end{theorem}

Since we have the semiclassical approximation $ {\cal F} ( H, \rho_t) = -i [ H_a + u H_b, \rho_t ]$ with $ u$ a control function, the SME \eqref{eq:SME-Ito1} belongs to the class of stochastic affine in control nonlinear differential systems, for which a number of stabilizability conditions have been developed \cite{Florchinger2,Florchinger1,Florchinger3}.
Call $ {\cal L}_0 $ the infinitesimal generator of the uncontrolled part of the dynamics
\[
{\cal L}_0  = {\cal L}  - {\cal L}_b u ,
\]
where
\[
{\cal L}_b \, \cdot = - \frac{i}{2} \left( [ H_b , \rho_t ] \frac{\partial \, \cdot }{ \partial \rho }  +  \frac{\partial \, \cdot }{ \partial \rho } [ H_b , \rho_t ] \right) .
\]

\begin{definition}
\label{def:stoch-Lyap-cond}
The SME \eqref{eq:SME-Ito1} satisfies a {\em stochastic Lyapunov condition} at $ \rho_d $ if $ \exists \; \mathcal{B}_\mathcal{M} \subset \mathcal{M}$ and a Lyapunov function $ V \in C^2 (\mathcal{B}_\mathcal{M} , \, \mathbb{R}) $ such that for all $ \rho_t \in \mathcal{B}_\mathcal{M} \smallsetminus \{ \rho_d\} $ for which the Lie derivative $ {\cal L}_b V_t =0 $ it is $ {\cal L}_0 V_t  < 0$.
The stochastic Lyapunov condition is almost global if $ \mathcal{B}_\mathcal{M} $ is all of $\mathcal{M}$ except for at most a finite number of isolated points.
\end{definition}

When this condition is fulfilled, $ V $ is said to be a stochastic control Lyapunov function for \eqref{eq:SME-Ito1}.
Our feedback synthesis relies on this condition, but does not follow any of the standard constructions for control Lyapunov functions \cite{Florchinger1}.

We shall instead make use of the following Jurjevic-Quinn type of stochastic stabilizability condition (see \cite{Florchinger1} Def. 3.5 and \cite{Florchinger2} Def. 3.1).

\begin{theorem}
\label{thm-stoch-Jurd-Quinn}
Assume $ \exists $ $ \mathcal{B}_\mathcal{M} \subset \mathcal{M} $, $ \rho_d \in \mathcal{B}_\mathcal{M} $ and $ V \in C^2 ( \mathcal{B}_\mathcal{M} , \, \mathbb{R} ) $, $ V ( \rho_d ) =0 $, $ V ( \mathcal{B}_\mathcal{M} \smallsetminus \{ \rho_d \} ) > 0$, such that
\begin{enumerate}
\item $ {\cal L} V_t  \leqslant 0 $ $ \forall \; \rho_t \in \mathcal{B}_\mathcal{M} $;
\item the set $ \left\{ \rho_t \in  \mathcal{B}_\mathcal{M} \text{ s. t. } {\cal L} ^{r+1 } V_t =  {\cal L} ^{r }  {\cal L}_b  V _t =0, \;  r \in \mathbb{N} \right\} = \left\{ \rho_d \right\} $.
\end{enumerate}
Then the feedback $ u_t = -  {\cal L}_b  V_t $ renders the equilibrium solution $ \rho_d $ locally asymptotically stable in probability.
\end{theorem}
All definitions and theorems carry on unchanged when $\mathcal{M} $ and \eqref{eq:SME-Ito1} are replaced by $ \mathcal{S} $ and \eqref{eq:SSE}.

\section{Continuous state reduction: the feedback stabilization problem}

The problem we will discuss and solve can be stated as follows.
\begin{prob}
Find a smooth control law $u(t)$ that (almost) globally stabilizes in
probability the pure state $\rho_d=\ket{\psi_d}\bra{\psi_d}$ of an
$N$-dimensional quantum system, whose dynamic is described by the
filtering equation \eqref{eq:SME-Ito1} conditioned by the
continuous observation of an observable $C$.
\end{prob}

We shall propose several choices of $u(t)$ as linear and nonlinear
feedback laws based on the conditional estimate for the state
$\rho_t$ at time $t$. In the physics literature, this approach has been
baptized bayesian feedback \cite{wiseman-bayesian} and of course
requires the real time integration of \eqref{eq:SME-Ito1}.
Since $ u(t) = u_t $ is smooth and adapted to the filtration $ {\cal E}_t $,
the closed loop solution exists and is unique in a global sense.

The first feedback law proposed (\S~\ref{sec:lin-feedb}) is linear and allows to achieve only local stabilizzability for the SSE (\S~\ref{sec:lin-feedb-SSE}).
If we choose a Lyapunov function that includes the variance of the measurement (\S~\ref{sec:var-Lyap}), then two modifications of the linear law are easily identifiable and are presented in \S~\ref{sec:nonlin-sum-of-squares} and \S~\ref{sec:nonlin-square-of-sum}.
The first one yields almost global asymptotic stability but only for the SSE, while with the second one we achieve almost global asymptotic stability in $ \mathcal{M}$ for the SME.
The relation between rate of convergence and gain tuning for the latter feedback is discussed in \S~\ref{sec:gain-tun}.

Let us first make suitable assumptions on $ {\cal F}$ and $ C$. In
order for $ \ket{\psi_d} $ to be an equilibrium, assume $
\ket{\psi_d }$  is an eigenstate of $ H_a $ and of $ C$. To avoid
unnecessary complications, assume further that the spectrum of $C$
is non-degenerate and that $ [ H_a, \, C]=0$. With this choice, it
is always possible for example to fix a basis such that $ \rho_d
=\ket{\psi_d} \bra{\psi_d} $ is diagonal and so are the free
Hamiltonian $ H_a $ and $ C$. We want to choose $ u $ so that $
\rho_d $ is rendered an attractor for the SME. Since the spectrum
of $ C$ is non-degenerate, $ \exists $ $N-1 $ state vectors other
than $ \ket{\psi_d} $, $ \ket{\psi_j} $, $ j =1, \ldots, N-1 $,
that are eigenstates of $ C$. When $ C $ is diagonal, they
correspond to diagonal density matrices
$\rho_j=\ket{\psi_j}\bra{\psi_j}$ with diagonal elements $ \{ 1,
0, \ldots , 0 \} $. Following the terminology of
\cite{Cla-qu-ens-feeb1}, we shall call these \emph{antipodal
states} of $ \rho_d $. Denote with $ \mathcal{J} $ the union of such
antipodal points: $ \mathcal{J} = \bigcup_{j=1}^{N-1}  \{ \ket{\psi_j}
\bra{\psi_j} \} $. Finally, to avoid trivial cases, assume that $
{\rm Graph} (H_b) $ is connected, i.e., that all transitions between
energy levels are enabled by the control field.

\subsection{A linear feedback controller}
\label{sec:lin-feedb}
A natural choice for a Lyapunov function is the distance between
density operators induced by the Hilbert-Schmidt norm
\cite{Nielsen1}:
\begin{equation}
V_1 = \tr ( \rho_d^2 ) - \tr( \rho_d \rho).
\label{eq:V-linear}
\end{equation}
One clearly sees that in the stochastic differential
\eqref{eq:inf-gen-Ito} the quadratic part can be neglected since $
V_1 $ is linear in $\rho_t$:
\beq
\begin{split}
{\cal L} V_{1, t}  & = - \tr ((-i[H,\rho_t] + {\cal D} ( C, \rho_t )) \rho_d) \\
& = - \tr ( -i[H_b,\rho_t] \rho_d) u = {\cal L}_b V_{1,t} u .
\end{split}
\label{eq:LV-linear-u}
\eeq
The non-Hamiltonian part vanishes because $C$ and $\rho_d$ commute
and the cyclic property of trace holds (see the proof of
Proposition~\ref{prop:diag}): $ \tr ( -i[H_a,\rho_t] \rho_d) =  0
$ and
\begin{equation}
\tr \left( \left( C\rho_t C-\frac{1}{2}(C^2\rho_t+\rho_t C^2 )\right) \rho_d \right) = \tr
\left(  (C^2\rho_t -C^2\rho_t) \rho_d \right) = 0. \label{eq:LV-linear}
\end{equation}
Hence in the SME \eqref{eq:SME-Ito1} this stabilization design is concerned
 only with the unitary part of the evolution and has the natural solution
\begin{equation}
u_t = k \, \tr ( -i[H_b,\rho_t] \rho_d), \quad k> 0.
\label{eq:feedb-control}
\end{equation}
Since the closed loop system has
\[
{\cal L}V_{1,t} = - k \, \tr^2 ( -i[H_b,\rho_t] \rho_d) \leqslant 0 ,
\]
one needs to study the $ \omega$-limit set of \eqref{eq:SME-Ito1} with the feedback \eqref{eq:feedb-control}.
This is the difficult part of the linear feedback design \eqref{eq:feedb-control}.
We certainly have the following for the set of (pure or mixed) diagonal density operators, call it $ {\cal Q}$ (often call eigenensemble \cite{Zyczkowski2}).

\begin{proposition}
\label{prop:diag} Consider the SME \eqref{eq:SME-Ito1}. For $
\rho_t \in{\cal Q} $, the state dynamics are not influenced by the
feedback \eqref{eq:feedb-control}. Moreover, ${\cal Q}$ is
invariant.
\end{proposition}

\proof It suffices to notice that:
\beq
\tr ( -i[H_b,\rho_t] \rho_d) =  0 \quad\forall\rho_t\textrm{
such that }[\rho_t,\rho_d]=0,
\label{eq:comm-diag}
\eeq
since $$\tr ( -i[H_b,\rho_t]
\rho_d) = \tr ( -i[H_b,\rho_t \rho_d]) = 0,$$ as any
commutator under the trace operation.
Similarly, $ [\rho_t,C] = [\rho_t, C^2]=0 $.
Thus, the only term affecting the dynamics is the diffusion term, which is also diagonal. Hence the diagonal set is invariant.
\qed

Since $ {\cal Q} $ is a convex set, we have the following.
\begin{corollary}
\label{cor:diag}
For the system \eqref{eq:SME-Ito1} with the feedback \eqref{eq:feedb-control} $ \nexists $ open neighborhoods $ \mathcal{B} \in {\rm Herm} $ such that $ \rho_d $ is locally asymptotically stable in probability in $ \mathcal{B}_\mathcal{M} = \mathcal{B} \cap \mathcal{M} $.
\end{corollary}

In fact, the dynamics confined to $ {\cal Q} $ is only a
fluctuation and since the probability of collapse to the
eigenstate $ \rho_a = \ket{\psi_j } \bra{\psi_j} \in \mathcal{J} $ is equal
to $ P _{\rho_a} = \tr(\rho_a \rho)$, it is never 1 if $ \rho\in {\cal Q}\smallsetminus \mathcal{J}$.

\subsection{Local stabilization of a class of Stochastic Schr\"odinger equations}
\label{sec:lin-feedb-SSE}
Consider the case $\eta=1$.
As discussed in Section~\ref{sec:model-SME-pres}, the SME \eqref{eq:SME-Ito1} is equivalent to the SSE \eqref{eq:SSE} and the state space is $ {\cal S}$.
From the transversality of $ {\cal S} $ with respect to the set of diagonal Hermitian matrices (see Theorem~E.2 of \cite{Frankel1}), the intersection of
$ {\cal S} $ with $ {\cal Q}$ is just $ \mathcal{J} \cup \{ \rho_d \} $.

For this relevant particular case, one can show
the following.

\begin{theorem}
\label{thm:SSE-loc-lin}
Assume $ \eta=1$ and that the following Kalman-like rank condition is satisfied:
\beq
{\rm rank } ( -i [ H_b ,\,  \rho_d ] ,\, [ A, \,  -i [ H_b ,\,  \rho_d ]] , \ldots , \underbrace{  \; [ \, A\; , \;\ldots \; , \, [\, A\; }_{\text{$N^2-N-1$ times}} , \,   -i [ H_b ,\,  \rho_d ]]\ldots ] ) = N^2-N
\label{eq:Kalman-like-in-theorem}
\eeq
where $ A$ is either $ -i H_a $ or $ C $.
Then the feedback law \eqref{eq:feedb-control} renders the equilibrium solution of \eqref{eq:SME-Ito1} locally asymptotically stable in probability.
\end{theorem}

\proof
In order to prove Theorem~\ref{thm:SSE-loc-lin}, we need a related deterministic result.
Consider the deterministic unitary bilinear control system obtained from \eqref{eq:SME-Ito1} in correspondence of $ C=0 $
\beq
\dot \rho = -i [ H_a , \, \rho ] - i \, u \, [ H_b , \,\rho ] , \qquad \rho \in {\cal S },
\label{eq:det-unit-bilin}
\eeq
and its tangent linear system at $ \rho_d $
\beq
\dot \rho = -i [ H_a , \, \rho ] - i \, u \, [ H_b , \,\rho_d ] .
\label{eq:det-unit-lin}
\eeq
\begin{lemma}
\label{lemma:det-Jurd-Q}
Assume $ H_a $ strongly regular. If \eqref{eq:det-unit-lin} satisfies the Kalman rank condition \eqref{eq:Kalman-like-in-theorem} with $ A = -i H_a $, then $ \rho_d $ is locally asymptotically stabilizable by means of the feedback \eqref{eq:feedb-control}.
\end{lemma}
Recall that $ A= A^\dagger $ strongly regular means $ A $ nondegenerate and with all transition frequencies (i.e., all differences of eigenvalues) that are different.
The proof of this Lemma is available in \cite{Cla-qu-ens-feeb1} (see also \cite{Mirrahimi2}).
It essentially relies on the Jurdjevic-Quinn condition \cite{Jurdjevic5}: starting from the identity
\[
\begin{split}
u & =  \tr ( -i[H_b,\, \rho] \rho_d)=0 \\
& = - \tr ( -i[H_b,\, \rho_d] \rho)=0 ,
\end{split}
\]
compute sufficiently many derivatives
\begin{subequations}
\label{eq:det-u=0}
\beqa
&  (-1)^2 \tr ( -i[H_a,\, -i[H_b,\, \rho_d]] \rho)=0 & \label{eq:det-u=0-A} \\
& (-1)^3 \tr ( -i[H_a ,\, -i[H_a,\, -i[H_b,\, \rho_d]]] \rho)=0 & \\ \label{eq:det-u=0-B}
& \vdots & \nonumber \\
&  (-1)^{r+1} \tr (\, \underbrace{-i[H_a ,\ldots , \, -i[H_a }_{\text{$ r$ times}},\, -i[H_b,\, \rho_d]]\ldots ] \rho)=0. &  \label{eq:det-u=0-C}
\eeqa
\end{subequations}
The strong regularity condition of $ H_a $ guarantees that all the commutators in \eqref{eq:det-u=0} are linearly independent up to a number equal to $ {\rm dim } ({\cal S}) $, i.e., \eqref{eq:Kalman-like-in-theorem} holds with $ A=-i H_a $, implying the controllability of the linearization \eqref{eq:det-unit-lin} and thus the local stabilizability of the original system \eqref{eq:det-unit-bilin}.
Since $ {\cal S} $ is a manifold (and not an Euclidean space like in \cite{Jurdjevic5}), the condition is only local.

Turning to the stochastic system \eqref{eq:SME-Ito1} and the stochastic Jurdjevic-Quinn condition of Theorem~\ref{thm-stoch-Jurd-Quinn}, since we have a linear Lyapunov function, when computing $ {\cal L}_0 {\cal L}_b V_{1,t} $ in $ u_t=0 $ the quadratic part of $ {\cal L}_0 $ does not appear:
\beq
\begin{split}
 {\cal L}_0 {\cal L}_b V_{1,t} & = \tr \left(  -i [H_b, \,{\cal F} (H_a , \rho_t)
+ {\cal D} (C, \rho_t ) ] \rho_d \right)  \\
& =  \tr \left(  -i [H_b,\, -i[H_a,\, \rho_t]]\rho_d \right)
+ \tr \left( -i [H_b,\, -\frac{\mu}{2} [C,\, [ C, \,  \rho_t ]]]  \rho_d \right) \\
& =  (-1)^2 \tr \left(  -i [H_a,\, -i[H_b,\, \rho_d]]\rho_t \right)
+ (-1)^3 \tr \left(  -\frac{\mu}{2} [C,\, [ C, \,-i [H_b, \,  \rho_d ]]]  \rho_t \right).
\label{eq:stoch-u=0-A}
\end{split}
\eeq
Similarly,
\beq
\begin{split}
 {\cal L}_0^2 {\cal L}_b V_{1,t}
 = & (-1)^3  \tr \left(  -i [H_a,\, -i[H_a,\,-i[H_b,\, \rho_d]]]\rho_t \right) \\
& +(-1)^4  \tr \left( -i[H_a,\, -\frac{\mu}{2} [C,\, [ C, \,-i [H_b,\,  \rho_d ]]]]  \rho_t \right) \\
& + (-1)^4 \tr \left(  -\frac{\mu}{2} [C,\, [ C ,\, -i[H_a,\, -i [H_b,\, \rho_d]]]] \rho_t \right)\\
& +(-1)^5 \tr \left(  -\frac{\mu}{2} [C,\, [ C,\, -\frac{\mu}{2} [C,\, [ C, \,-i [H_b,\,  \rho_d ]]]]]  \rho_t \right)
\label{eq:stoch-u=0-B}
\end{split}
\eeq
and so on for $  {\cal L}_0^r {\cal L}_b V_{1,t} $, $ r>2$.
Hence, in the case of $H_a $ strongly regular the stochastic Jurdjevic-Quinn condition holds whenever Lemma~\ref{lemma:det-Jurd-Q} holds, as the Lie algebra spanned by the commutators in $ {\cal L}_0^r {\cal L}_b V_1  =0 $ is at least as large as the one spanned by the commutators appearing in the deterministic conditions $ \frac{d^r u} {d t^r} =0 $ \footnote{Notice that since $ V_1 $ is linear, in Theorem~\ref{thm:SSE-loc-lin} we are only concerned with the linear part of the infinitesimal generators and this allows to infer the stochastic Jurdjevic-Quinn condition directly in terms of the Lie algebra, just like in its deterministic counterpart. Redoing the computations above for the corresponding Stratonovich equation (for which Lie algebraic conditions can be made fully rigorous for any $ V_1 $)
\[
d\rho_t  = ({\cal F}( H,\rho_t) + {\cal D} ( C, \rho_t ) -\frac{1}{2} {\cal G}_s (C, \rho_t) ) dt+  {\cal G} (C, \rho_t)\circ d W_t ,
\]
where the quadratic term in the drift, $ {\cal G}_s (C, \rho_t) $, is (in the case $ C$ is traceless)
\[
\begin{split}
 {\cal G}_s (C, \rho_t) & = \frac{1}{2} \left(  {\cal G} (C, \rho_t) \frac{\partial {\cal G} (C, \rho_t) } {\partial \rho} + \frac{\partial {\cal G} (C, \rho_t) } {\partial \rho} {\cal G} (C, \rho_t) \right) \\
& = \mu \eta \left( 2 C \rho_t C +  C^2 \rho_t + \rho_t C^2  - 4 \langle C \rangle_t ( C \rho_t + \rho_t C ) + 4 \langle C \rangle_t^2 \rho_t \right) ,
\end{split}
\]
one arrives at the same conclusion.}.
However, even when $ H_a $ not strongly regular but $ C $ is, the stochastic Jurdjevic-Quinn condition still holds as the terms
\[
  -\frac{\mu}{2} [C,\, [ C,\,\ldots , -\frac{\mu}{2} [C,\, [ C, \,-i [H_b,\,  \rho_d ]]]\ldots ]]
\]
still provide the needed linearly independent commutators (see \cite{Cla-contr-root1} for explicit recursive computations of the commutators involved).
Since this is implied by \eqref{eq:Kalman-like-in-theorem}, the proof is completed.
The condition is local just like its deterministic counterpart.
\qed

Notice that the feedback law \eqref{eq:feedb-control} rewritten
for the SSE \eqref{eq:SSE} is \beq
\begin{split}
u_t &=k\tr(-i[H_b,\,\ket{\psi_t}\bra{\psi_t}]\ket{\psi_d}\bra{\psi_d})\\
&=-ik(\langle\psi_t|\psi_d\rangle\langle\psi_d|H_b|\psi_t\rangle-\langle\psi_t|\psi_d\rangle^*\langle\psi_d|H_b|\psi_t\rangle^*)\\
&=-2k\textrm{Im}(\langle\psi_t|\psi_d\rangle\langle\psi_d|H_b|\psi_t\rangle).
\end{split}
\eeq

\begin{remark}
\label{rem:H_b-has-all-erms}
Assume, without loss of generality, that $ \rho_d = {\rm diag} \{ 1, 0, \ldots, 0 \}$.
Then in order for \eqref{eq:Kalman-like-in-theorem} to hold it must be $ ( H_b)_{1j} = (H_b)_{j1}^* \neq 0 $, i.e., the control Hamiltonian $H_b $ must ``enable'' all transitions from $ \ket{\psi_d} $ to all other eigenstates $ \ket{\psi_j } $, $ j = 1, \ldots, N-1$.
\end{remark}

When $ \eta<1 $, the structure of the state space is larger than $ {\cal S} $ and in particular the transversality of the state space with respect to $ \mathcal{Q}$ no longer holds, hence Theorem~\ref{thm:SSE-loc-lin} does not apply.

\subsection{A variance-based Lyapunov condition}
\label{sec:var-Lyap}
The feedback \eqref{eq:feedb-control} is the same linear
controller used in \cite{Cla-qu-ens-feeb1} to study the
deterministic stabilization problem with state feedback
(corresponding to $C=0$). In that setting, its region of
attraction does not correspond to the entire state space. In our
stochastic problem, there is the additional requirement that the $
\omega$-limit set has to be invariant also to the flow of the
diffusion part. We will exploit this feature, considering, instead
of $ V_1$, the following candidate Lyapunov function:
\begin{equation}
V = V_{1} + V_{2} = \tr ( \rho_d^2 ) - \tr( \rho_d \rho) +
\tr ( C^2 \rho) - \tr^2 ( C \rho). \label{eq:V-nonlinear}
\end{equation}
Clearly $ V\geqslant 0$ , $ V=0 $ only in $ \rho= \rho_d $.
The function $ V_2 $
in \eqref{eq:V-nonlinear} is the variance of the filtering process
along $ C$:
\begin{equation}
V_{2} = \langle C^2\rangle-\langle C\rangle^2.
\label{eq:V-variance}
\end{equation}
$ V_{2} $ has the property of being a positive semidefinite Morse function on $\mathcal{M}$, i.e., a function whose critical points are nondegenerate \cite{Guillemin1} and can be used to attain a stochastic Lyapunov condition.

\begin{theorem}
\label{thm:SME-stoch-Lyap-cond}
The system \eqref{eq:SME-Ito1} satisfies an almost global stochastic Lyapunov condition with respect to the Lyapunov function $V$ given in \eqref{eq:V-nonlinear}.
The only points for which the stochastic Lyapunov condition is not satisfied are the $N-1 $ antipodal states of $ \rho_d $.
\end{theorem}

\proof To discuss the asymptotic properties of $ V_{2} $, it is
useful to notice first that, using $[H,C]=0,$ the cyclic property
of the trace and Ito's rule: \beq
\begin{split}
d\langle C \rangle_t&=\tr(Cd\rho_t)\\
&=\tr ( -i [ H , \, \rho_t ]C) dt + 2\sqrt{\mu \eta }\tr(C(C-\langle C \rangle_t)\rho_t)dW_t\\
&=u \tr ( -i [ H_b , \, \rho_t ]C) dt + 2\sqrt{\mu \eta }(\tr(C^2\rho_t)-\langle C\rangle^2_t))dW_t\\
&=u \tr ( -i [ H_b , \, \rho_t ]C) dt + 2\sqrt{\mu \eta
}V_{2,t}dW_t.
\end{split}
\eeq
Thus, for the system \eqref{eq:SME-Ito1} we have:
\begin{equation}
\label{dV}
\begin{split}
dV_{2,t} &= d\langle C^2\rangle_t-2 \langle C\rangle_td \langle C \rangle_t- (d\langle C \rangle _t)^2\\
&= (\tr ( -i [H,\, \rho_t] ( C^2 - 2 \langle C \rangle_t C ) ) -4\mu \eta V^2_{2,t}) dt \\
& +2\sqrt{\mu \eta }(\tr(C^3\rho_t-\langle C\rangle_t C^2\rho_t) -2 \tr(\langle C\rangle_t C^2\rho_t)+2 \langle C \rangle^3_t)dW_t\\
&= (u \; \tr ( -i [H_b,\, \rho_t] ( C^2 - 2\langle C \rangle_t C )
) - 4\mu \eta V^2_{2, t} ) dt+ 2 \sqrt{\mu \eta }
\sigma(C,\rho_t)dW_t,
\end{split}
\eeq with $\sigma(C,\rho_t)=\langle C^3\rangle_t-3 \langle C
\rangle_t \langle C^2\rangle _t+2\langle C \rangle^3_t$ the 3rd
central moment.

From \eqref{eq:V-nonlinear}, \eqref{dV} and \eqref{eq:LV-linear-u}, the stochastic differential \`a la It\^o for $ V$ is
\beq
{\cal L} V_t  =  - tr ( -i[H_b,\rho_t] (\rho_d + 2 \langle C \rangle_t C - C^2) ) u - 4 \mu \eta \left( \langle C^2 \rangle_t  - \langle C \rangle_t^2) \right)^2.
\label{eq:LV-sum-12}
\eeq
Considering the zero-control behavior, notice that if $V_{2,t}=0$ the system state must be in an eigenstate of $ C$ and hence $\sigma(C,\rho_t)=0$.
Thus $V_{2,t} =0$ is stationary for \eqref{dV}.
The convergence to the eigenstates follows applying Theorem~\ref{thm-Lyap-stab1} to any bounded right interval of zero containing $V_{2,0} \neq 0 $. In fact,
\[
\left. {\mathcal L} V_{2,t} \right|_{u=0} =-4\mu \eta V^2_{2,t}<0
\]
proves the convergence in probability of the variance to zero.
Hence, for the closed loop system, looking at \eqref{eq:LV-sum-12}, $ {\cal L}_0 V_t = \left. {\cal L}V_{2,t}\right|_{u=0} < 0 $ everywhere, except at the $N-1 $ other eigenvalues of $ C$ and Definition~\ref{def:stoch-Lyap-cond} applies almost globally.
\qed

\subsection{Almost global stabilization of the SSE by nonlinear feedback}
\label{sec:nonlin-sum-of-squares}
\begin{theorem}
\label{thm:SSE-almost-glob-nonlin}
Assume $ \eta=1$. The feedback law
\beq
u_t = k \, \tr ( -i [H_b,\rho_t] (\rho_d + 2 \langle C \rangle_t C - C^2) ), \quad k>0,
\label{eq:feedb-non-sum-of-squares}
\eeq
renders the equilibrium solution $ \rho_d $ of the SSE \eqref{eq:SME-Ito1} almost globally asymptotically stable in probability, with region of attraction given by $ {\cal S} \smallsetminus \mathcal{J}$.
\end{theorem}
\proof
The feedback \eqref{eq:feedb-non-sum-of-squares} makes \eqref{eq:LV-sum-12} into a negative semidefinite sum of squares:
\beq
{\cal L} V_t  =  - k tr^2 ( -i[H_b,\rho_t] (\rho_d + 2 \langle C \rangle_t C - C^2) )  - 4 \mu \eta \left( \langle C^2 \rangle_t  - \langle C \rangle_t^2 \right)^2\leqslant 0.
\label{eq:LV-sum-closed-loop-12}
\eeq
Calling $ {\cal N}_\mathcal{S} $ the set of critical points of $ V$: $ {\cal N}_\mathcal{S} = \{ \rho_t \in {\cal S} \text{ s. t. } {\cal L} V_t  =0 \} $, from Theorem~\ref{thm:LaSalle} we need to compute the $ \omega$-limit set of the closed loop inside ${\cal N}_\mathcal{S}$.
Since \eqref{eq:LV-sum-closed-loop-12} is a sum of squares, $ {\cal N}_\mathcal{S}$ must be a subset of $ \{ \rho_t \in {\cal S}  \text{ s. t. } \left. {\cal L} V_{2,t}  \right|_{u=0}  =0 \} = \mathcal{J} $.
Since in $ \rho_a\in \mathcal{J}$, $ \rho_a\neq  \rho_d $, $ V( \rho_a ) > 0$, $ \rho_a $ cannot be asymptotically stable in probability.
\qed

\begin{remark}
Notice that unlike Theorem~\ref{thm:SSE-loc-lin}, Theorem~\ref{thm:SSE-almost-glob-nonlin} does not require any special structure for $H_b $ (compare Remark
following Theorem~\ref{thm:SSE-loc-lin}).
In loose terms, one could say that while the design of Theorem~\ref{thm:SSE-loc-lin} relies on a controllable linearization, in Theorem~\ref{thm:SSE-almost-glob-nonlin} uncontrollable, asymptotically stable modes are allowed in the linearization.
\end{remark}

Of course, it can be easily shown that the feedback \eqref{eq:feedb-non-sum-of-squares} can be used also in place of \eqref{eq:feedb-control} in Theorem~\ref{thm:SSE-loc-lin}.

\subsection{Almost global stabilization of the SME by nonlinear feedback}
\label{sec:nonlin-square-of-sum}
When $ \eta \leqslant 1 $, Proposition~\ref{prop:diag} and Corollary~\ref{cor:diag} still hold also with the feedback \eqref{eq:feedb-non-sum-of-squares}. Hence $ \rho_t \in {\cal Q}$ is not attracted to $\rho_d $ with probability 1.
Although simulation results seem to suggest that with both the feedback laws \eqref{eq:feedb-control} and \eqref{eq:feedb-non-sum-of-squares} all non-diagonal density operators are attracted with probability 1 to $ \rho_d $, we do not see any clear way to prove it.
The problem can however be solved in full generality by a different choice of feedback.

\begin{theorem}
\label{thm:nonlin-feedb1} The system \eqref{eq:SME-Ito1} with
feedback law
\begin{equation}
u_t =  \tr ( -i[H_b,\rho_t] (\rho_d + 2 \langle C \rangle_t C - C^2)) - 4 \sqrt{\mu \eta} \left( \langle C^2 \rangle_t - \langle C \rangle_t ^2 \right)
\label{eq-feedb-nonlin1}
\end{equation}
admits $ \rho_d $ as equilibrium solution which is almost globally
asymptotically stable. The only states in $\mathcal{M}$ which are
not attracted in probability to $ \rho_d $ are its $N-1 $
antipodal states.
\end{theorem}

\proof

Consider still the Lyapunov function $ V$ given in \eqref{eq:V-nonlinear}.
It is easy to see that the nonlinear feedback \eqref{eq-feedb-nonlin1} completes $ {\cal L} V_t $ in \eqref{eq:LV-sum-12} to a square:
\[
{\cal L} V_t  =  - \left( \tr ( -i[H_b,\rho_t] (\rho_d + 2 \langle C \rangle_t C - C^2))  - 2  \sqrt{ \mu \eta} \left(  \langle C^2 \rangle_t - \langle C \rangle_t ^2 \right) \right)^2 \leqslant 0.
\]
Notice first that $ \rho_a \in \mathcal{J} $ is a stationary point of both open and closed loop systems: $ {\cal L}_0 V (\rho_a ) = {\cal L} V (\rho_a ) =0$.
We need to show that ${\cal N} $ cannot belong to the $ \omega$-limit set of the closed loop (with the exclusion of $ \mathcal{J}$) and that there is no subset of $ \mathcal{M} \smallsetminus \mathcal{J}$ which can remain undriven for $ t \to \infty $.
The crucial difference with respect to \eqref{eq:feedb-non-sum-of-squares} is that \eqref{eq-feedb-nonlin1} implies $ {\cal L}_0 V_t \neq \left. {\cal L} V_t \right|_{u=0 } $.
From Theorem~\ref{thm:SME-stoch-Lyap-cond}, the stochastic Lyapunov condition holds true and guarantees $ {\cal L}_0 V_t < 0 $ almost globally.
In particular, notice that $ {\cal L}_0 V_t < 0 $ everywhere in $ {\cal N} \smallsetminus \mathcal{J}$, hence $ {\cal N} \smallsetminus \mathcal{J} $ cannot belong to the $ \omega$-limit set.
In addition $ \left. u_t \right|_{\cal N} = -2 \sqrt{ \mu \eta } V_{2,t} <0 $, i.e., the set of spurious critical points is evaded also by means of the control action.
The other claim follow by the similar observation that the zero-feedback locus
\[
U = \{ \rho_t \in \mathcal{M} \text{ s. t. } \tr(  -i[H_b,\rho_t] (\rho_d + 2 \langle C \rangle_t C - C^2)) = 4 \sqrt{\mu \eta} \left( \langle C^2 \rangle_t - \langle C \rangle_t ^2   \right) \}
\]
is never invariant to $ {\cal L}_0 V_t $ outside $ \mathcal{J}$.
\qed

\begin{remark}
If the domain of attraction of \eqref{eq-feedb-nonlin1} is $ \mathcal{M} \smallsetminus \mathcal{J} $, the critical points of $ \mathcal{J} $ are automatically repulsive equilibria of the closed loop system.
If for some $ s\geqslant 0$, $ \rho_s \in \mathcal{J}$, then an arbitrarily small unitary open loop perturbation is enough to evade from $ \mathcal{J}$.
\end{remark}

\subsection{Gain tuning and rate of convergence}
\label{sec:gain-tun}
The performances of the feedback design \eqref{eq-feedb-nonlin1} can be improved by adding, and tuning appropriately, two different gains.
Instead of \eqref{eq-feedb-nonlin1}, consider the following
\begin{equation}
u_t = k^2 \, \tr ( -i[H_b,\rho_t] (\rho_d + (2 \langle C \rangle_t C - C^2)/\ell^2)) - 4 \frac{k \sqrt{\mu \eta}}{\ell} \left( \langle C^2 \rangle_t - \langle C \rangle_t ^2 \right),
\label{eq-feedb-nonlin2}
\end{equation}
with $ k>0$, $ \ell>0$.
While $ k $ corresponds to the usual feedback gain, $ \ell$ can be thought of as a rescaling of the variance in the Lyapunov function \eqref{eq:V-nonlinear}:
\[
\tilde{V} = V_1 + \frac{1}{\ell^2} V_2.
\]
The corresponding closed loop stochastic generator is then:
\beq
{\cal L} \tilde{V}_t  =  - \left( k \, \tr ( -i[H_b,\rho_t] (\rho_d + (2 \langle C \rangle_t C - C^2))/\ell^2)  - 2  \frac{\sqrt{ \mu \eta} }{\ell} \left(  \langle C^2 \rangle_t - \langle C \rangle_t ^2 \right) \right)^2 .
\label{eq:LV-nonlin-rescaled}
\eeq

\begin{proposition}
The equilibrium solution $ \rho_d $ of the system \eqref{eq:SME-Ito1} with the feedback law \eqref{eq-feedb-nonlin2} is almost globally asymptotically stable for all $ k> 0 $ and $ \ell > 0$.
When $ \ell \to \infty$ one recovers the linear feedback \eqref{eq:feedb-control}.
\end{proposition}
\proof
From \eqref{eq:LV-nonlin-rescaled}, for the closed loop system $ {\cal L} \tilde{V}_t < 0 $ in $ {\cal N} \smallsetminus \mathcal{J} $, $ \forall\;  k>0 $ and $ \forall $ finite $ \ell>0$, as the proof of Theorem~\ref{thm:nonlin-feedb1} still applies. The limit behavior follows by inspection of \eqref{eq-feedb-nonlin2}.
\qed

Notice that tuning $ \ell $ does not correspond to modulating the strength of the weak measurement $ \mu$.

The effect of $ \ell $ is to change the influence of $ V_2 $ on the closed loop dynamics.
In terms of the Hilbert-Schmidt norm $ V_1 $, from \eqref{eq:LV-linear-u}, its close loop differential is given by
\[
\begin{split}
& -  k^2 \, \tr^2 ( -i[H_b,\rho_t] \rho_d)  - \frac{k^2}{\ell^2}  \tr ( -i[H_b,\rho_t]  (2 \langle C \rangle_t C - C^2)) \tr( -i[H_b,\rho_t] \rho_d) \\
& +  4 \frac{k \sqrt{\mu \eta}}{\ell} \tr ( -i[H_b,\rho_t] \rho_d) V_{2,t}.
\end{split}
\]
Since only the first term is sign definite, convergence in probability to the target state $ \forall \; \rho_0 \in \mathcal{M} \smallsetminus \mathcal{J} $ is guaranteed to be faster if we raise the gain $ \ell$ (recall that $ V_{2} $ was introduced only to perturb the ``symmetry'' of the problem).

\section{Example: 2 level case}

For $N=2 $, it is possible to give a simple pictorial description of the trajectories of the system, provided one chooses a real parametrization, like the triple $( x, \, y, \, z ) $ representing the Bloch vector: $ \rho = \frac{1}{2} ( I_2 + x \sigma_x + y \sigma_y + z \sigma_z )$, with $\sigma_x, \, \sigma_y , \,\sigma_z$ the Pauli matrices.
Consider the observable $ C = \sigma_z $, its eigenstate $ \rho_d = \begin{bmatrix} 1 & 0 \\ 0 & 0 \end{bmatrix} $ and the Hamiltonian $ H_a = h_a \sigma_z $, $ H_b = \sigma_x $.
Since
\beqan
-i [ H_a , \, \rho_t] & = & h_a (- y_t \sigma_x + x_t \sigma_y ) , \\
-i [ H_b , \, \rho_t] & = & -z_t \sigma_y + y_t \sigma_z , \\
{\cal D} (C, \, \rho_t) & = & - \mu ( x_t \sigma_x + y_t \sigma_y ) , \\
{\cal G} ( C, \, \rho_t ) & = & \sqrt{ \mu \eta } ( - x_t z_t \sigma_x - y_t z_t \sigma_y + (1-z_t^2 ) \sigma_z ) ,
\eeqan
the SME \eqref{eq:SME-Ito1} in terms of the Bloch vector is
\beq
\begin{split}
d x_t & = ( -h_a y_t - \mu x_t ) dt - \sqrt{\mu \eta } x_t z_t d W_t \\
d y_t & = ( h_a x_t - u z_t - \mu y_t ) dt - \sqrt{\mu \eta } y_t z_t d W_t \\
d z_t & = u  y_t  dt - \sqrt{\mu \eta } ( 1- z_t^2) d W_t .
\end{split}
\label{eq:SME-Bloch}
\eeq
The Lyapunov functions are
\[
V_1 = \frac{1}{2} ( 1-z ) , \qquad V_2 = 1- z^2 ,
\]
hence
\[
{\cal L}\tilde{V}_t = - y_t \left( 1 - \frac{4 z_t}{\ell^2} \right) u - \frac{ 4 \mu \eta }{\ell^2 } ( 1-z_t^2 ) ^2 .
\]
The feedback laws \eqref{eq:feedb-control}, \eqref{eq:feedb-non-sum-of-squares} and \eqref{eq-feedb-nonlin1} are, respectively,
\beqa
u_t & = & k \, y_t \label{eq:feedb-control-Bloch}, \\
u_t & = & k \, y_t \left( 1 - \frac{4 z_t}{\ell^2} \right) , \label{eq:feedb-non-sum-of-squares-Bloch} \\
u_t & = & k^2  y_t \left( 1 - \frac{4 z_t}{\ell^2} \right)
- \frac{4 k \sqrt{\mu \eta } }{\ell} ( 1-z_t^2 ) ,
\label{eq-feedb-nonlin1-Bloch}
\eeqa
for some $ k>0 $, $\ell > 0$.
In correspondence of $ \tilde{V} $ and \eqref{eq:feedb-non-sum-of-squares-Bloch}, the closed loop infinitesimal generator is the sum of squares
\[
{\cal L}\tilde{V}_t = - k \left( y_t \left( 1 - \frac{4 z_t}{\ell^2} \right)\right) ^2 - \frac{ 4 \mu \eta }{\ell^2 } ( 1-z_t^2 ) ^2 ,
\]
while for \eqref{eq-feedb-nonlin1-Bloch} it is the square of a sum
\beq
{\cal L}\tilde{V}_t = - \left( k y_t \left( 1 - \frac{4 z_t}{\ell^2} \right) - \frac{ 2\sqrt{ \mu \eta }}{\ell } ( 1-z_t^2 ) \right)^2 .
\label{L-tilde-V-feedb-nonlin1-Bloch}
\eeq
Looking at the closed loop dynamics, one sees that on the ``line'' of diagonal densities $ x=y=0$, for \eqref{eq:feedb-control-Bloch} and \eqref{eq:feedb-non-sum-of-squares-Bloch} the feedback is $0$, the line itself is invariant and the dynamics driven only by the filtering term.
This is no longer true for \eqref{eq-feedb-nonlin1-Bloch}, as expected.

The level curves of \eqref{L-tilde-V-feedb-nonlin1-Bloch} are visualized in Fig~\ref{fig:L-tilde-V} for different choices of the parameters $ k$, $ \ell$. The figures shows that for $ \ell $ that grows the locus $ {\cal L}\tilde{V}_t =0 $ tends to become aligned with the axis $ y=0 $. The effect of raising $ k$ instead is to increase the rate of convergence.

\begin{figure}[ht]
\begin{center}
\includegraphics[width=6cm]{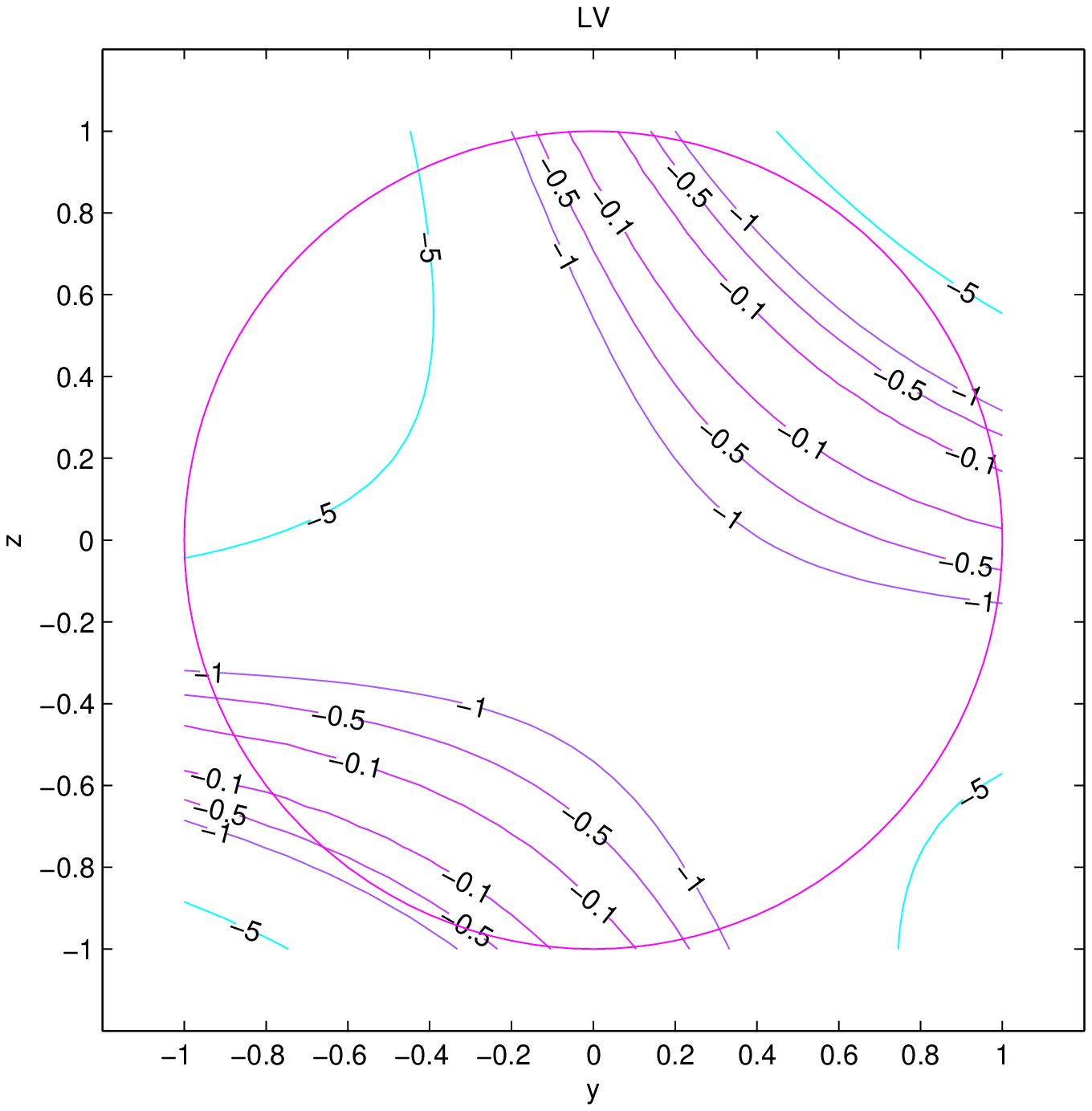}
\includegraphics[width=6cm]{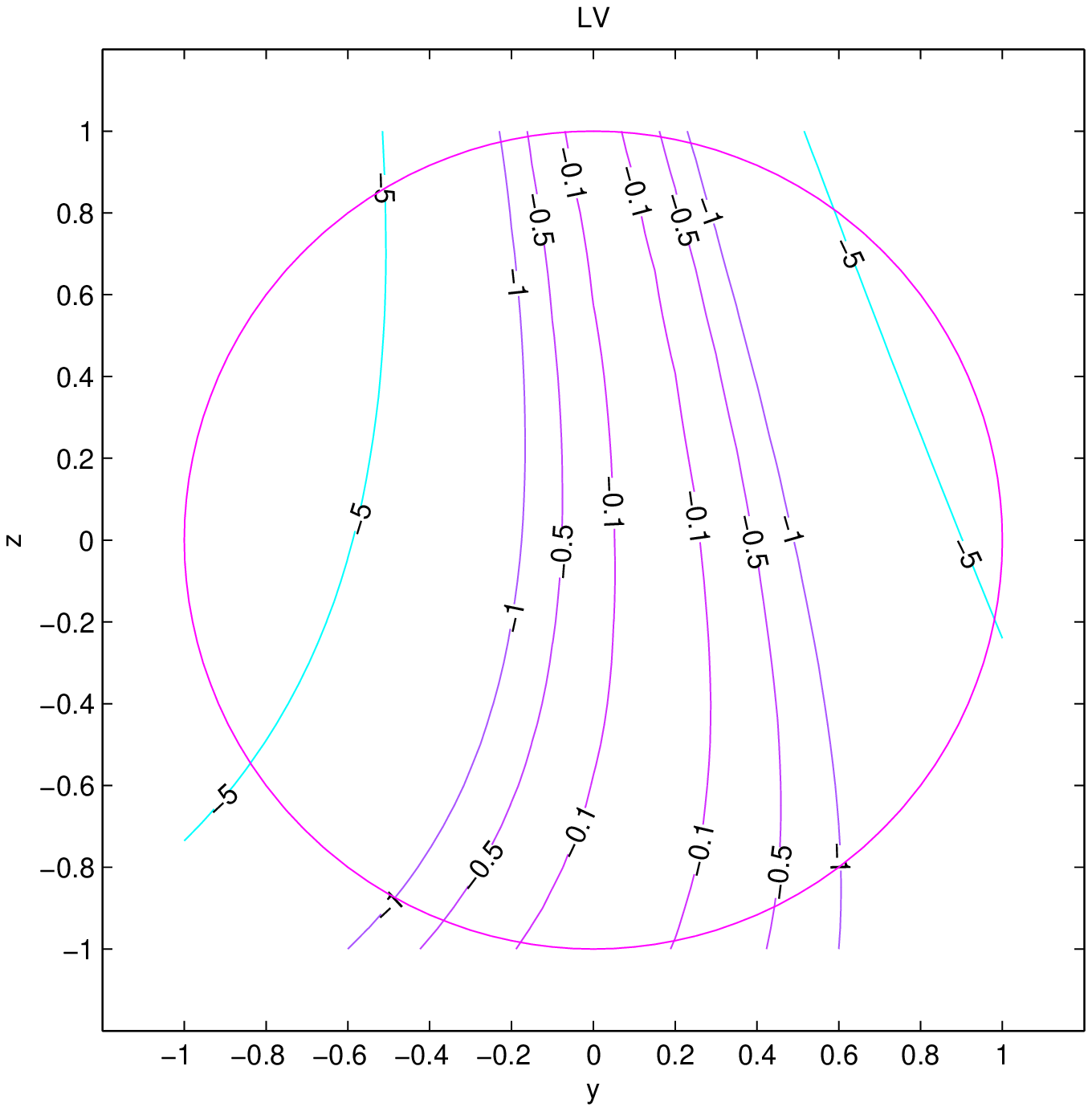}
\includegraphics[width=6cm]{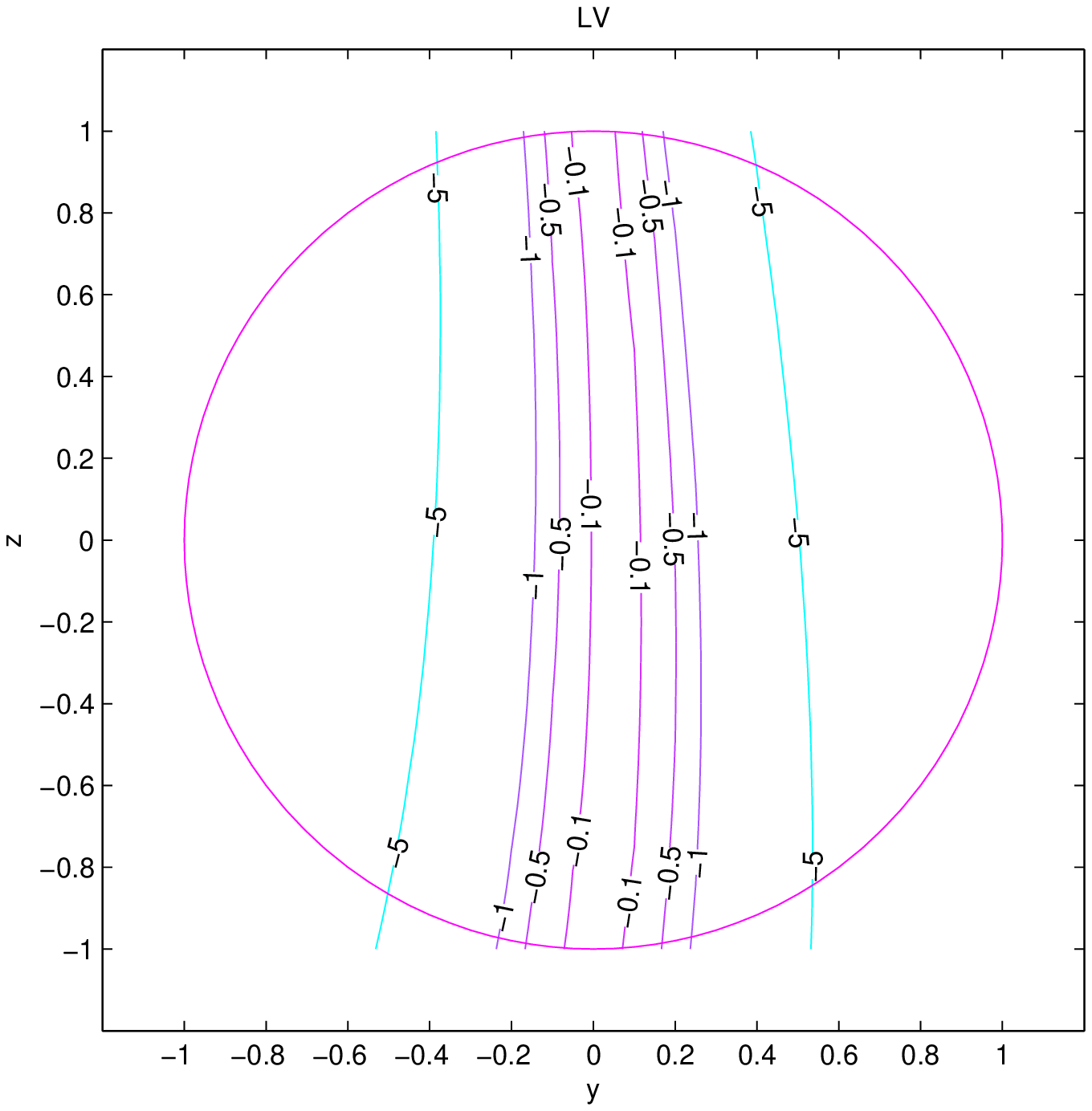}
 \caption{Level surfaces $ {\cal L}\tilde{V} =\text{const} $ (eq.\eqref{L-tilde-V-feedb-nonlin1-Bloch}) in the $ (y,z) $ plane for different values of $ k, \ell$, in correspondence of $ \mu=1$, $ \eta=1/2$. Top left: $(k,\ell )= ( 1,1)$; Top right: $(k,\ell )= ( 3,3)$; bottom: $(k,\ell )= ( 5,5)$.}
\label{fig:L-tilde-V}
\end{center}
\end{figure}

For the feedback \eqref{eq-feedb-nonlin1-Bloch}, in Fig.~\ref{fig:traj-2-lev-sphere} a few sample trajectories are plotted for different initial conditions (the boldface trajectory corresponds to $ \rho_0 = I_2/2$, i.e., to the maximally mixed state).
They are reproduced as $ x_t, \, y_t, \, z_t $ versus time in Fig.~\ref{fig:traj-2-lev-Bloch}.
The corresponding time courses of $ u_t $, $ \tr (\rho_t ^2 ) $, $ \tilde{V}_t $ and $ {\cal L}\tilde{V}_t $ are shown in Fig.~\ref{fig:u-pur-V-LV}.
\begin{figure}[ht]
\begin{center}
\includegraphics[width=9cm]{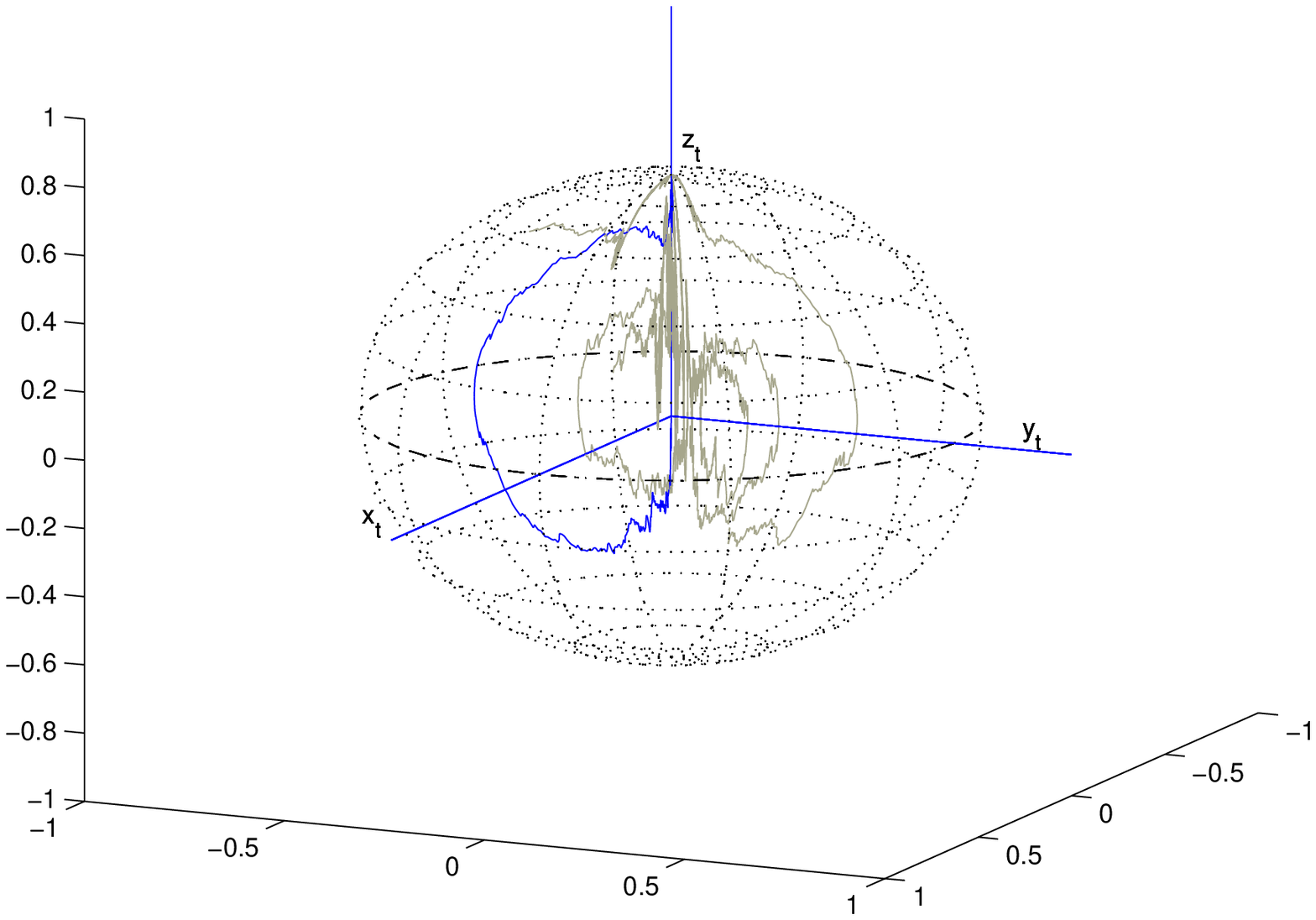}
 \caption{A few trajectories of the closed loop system \eqref{eq:SME-Bloch}-\eqref{eq-feedb-nonlin1-Bloch} for $ \mu=1$, $ \eta=1/2$. The boldface trajectory corresponds to $ \rho_0 = I_2/2$.}
\label{fig:traj-2-lev-sphere}
\end{center}
\end{figure}

\begin{figure}[ht]
\begin{center}
\includegraphics[width=9cm]{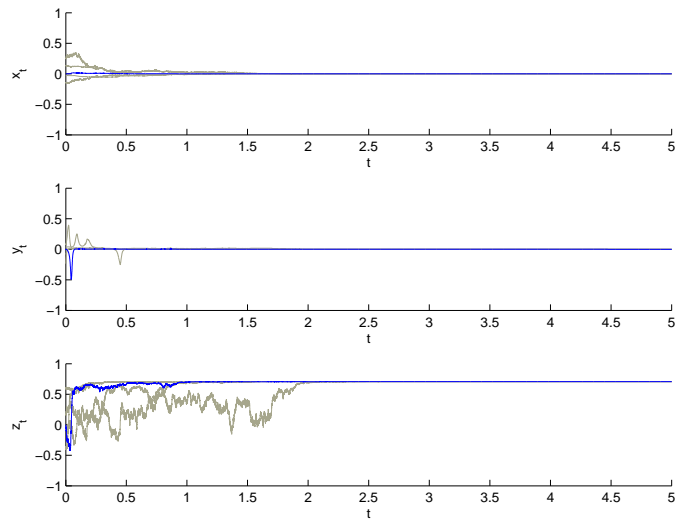}
 \caption{The same trajectories of Fig.~\ref{fig:traj-2-lev-sphere} versus time.}
\label{fig:traj-2-lev-Bloch}
\end{center}
\end{figure}

\begin{figure}[ht]
\begin{center}
\includegraphics[width=9cm]{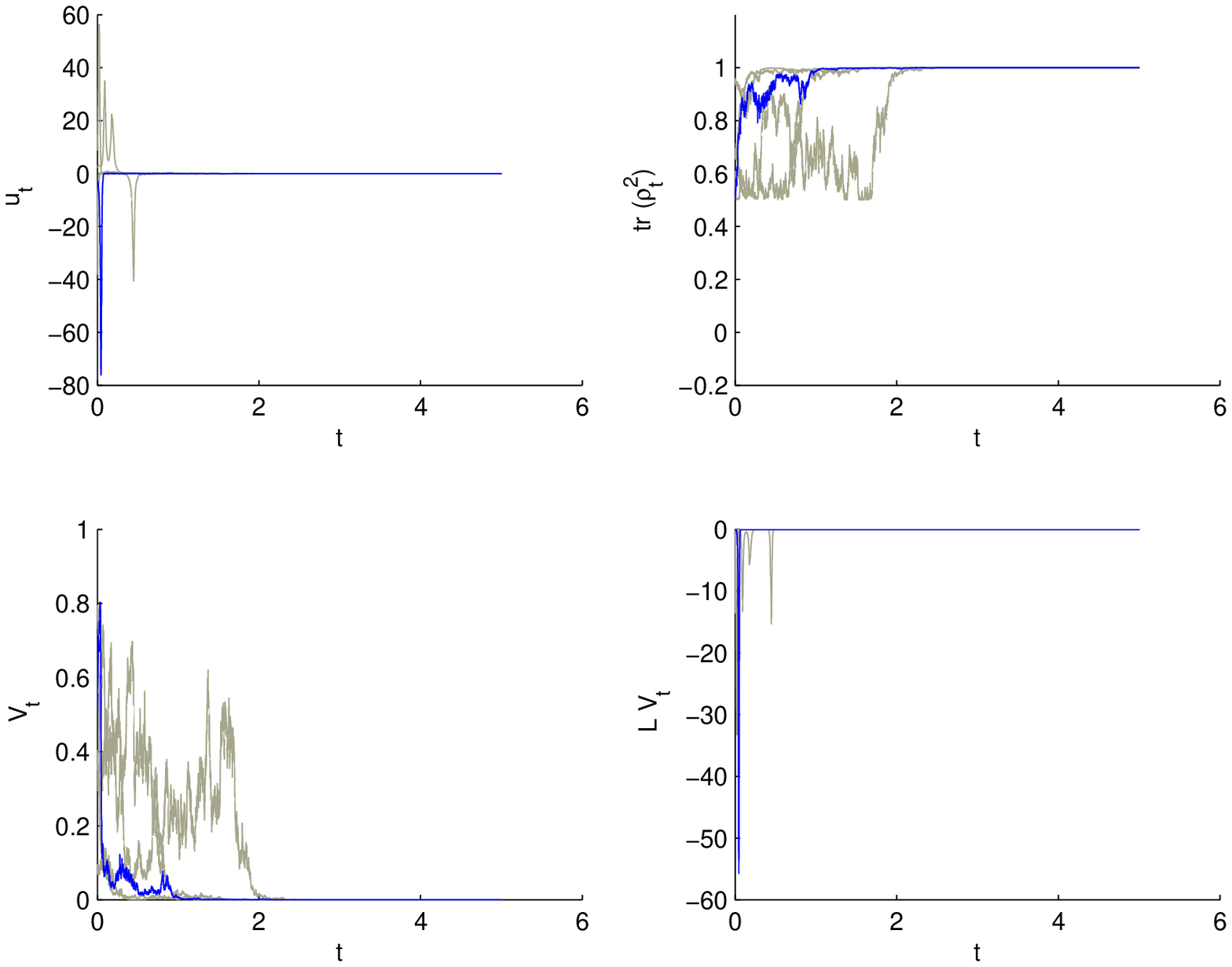}
 \caption{Top left: control signal $ u_t$; top right: $ \tr( \rho_t ^2 ) $; bottom left $ \tilde{V}_t $; bottom right: $ {\cal L}\tilde{V}_t $. The trajectories are the same as in Figg.~\ref{fig:traj-2-lev-sphere}-\ref{fig:traj-2-lev-Bloch}}
\label{fig:u-pur-V-LV}
\end{center}
\end{figure}

\section{Conclusion}
It is well known in Control Theory that finding a Lyapunov function for nonlinear systems is more an art than a systematic science, and that the knowledge about the physical process can provide the intuition necessary for this scope.
The present work is nothing but a confirm of both these rules of thumb.
We consider (some of) the standard design procedures available for the class of stochastic differential equations we deal with and show how they provide only a partial solution to our stabilization problem.
Once we integrate this design with some physical insight on the structure of the SDE, however, the feedback synthesis becomes much more efficient and allows for a simple analytic demonstration regardless of the dimension of the system.
In addition, since the Lyapunov function is a Morse function on the space of density operators, the feedback stabilization design guarantees global convergence up to a finite number of isolated and repulsive critical points.

\noindent{\bf Acknowledgements} The authors would like to thank R.
van Handel and S. Grivopoulos for useful discussions on the topic
of this work.

\bibliographystyle{plain}




\begin{thebibliography}{1}

\bibitem{Adler1}
S. L. Adler, D. C. Brody, T. A. Brun and L. P. Hughston.
\newblock Martingale models for quantum state reduction,
\newblock J. Phys. A: Math. Gen. {\bf 34} 8795-8820, 2001.

\bibitem{Cla-contr-root1}
C.~Altafini.
\newblock Controllability of quantum mechanical systems by root space
  decomposition of su({N}).
\newblock {\em Journal of Mathematical Physics}, {\bf43}:2051--2062, 2002.


\bibitem{Cla-qu-ens-feeb1}
C.~Altafini.
\newblock Feedback stabilization of quantum ensembles: a global convergence
  analysis on complex flag manifolds.
\newblock {\em Preprint arXiv:quant-ph/0506268}, 2004.


\bibitem{Arnold1}
L.~Arnold.
\newblock {\em Stochastic Differential Equations: Theory and Applications}.
\newblock Krieger Publishing Company, Malabar, Florida, 1974.

\bibitem{Battilotti1}
S.~Battilotti and A.~De~Santis.
\newblock Stabilization in probability of nonlinear stochastic systems with guaranteed cost.
\newblock SIAM J. Contr. Optim. {\bf 40}: 1938--1964, 2002.


\bibitem{belavkin-nondemolition}
V. P. Belavkin.
\newblock Nondemolition measurements and control in quantum dynamical systems.
\newblock In {\em Proceedings, Information Complexity and Control in Quantum
Physics, Udine 1985} (A. Blaquiere, S. Diner and G. Lochak Eds.).
\newblock pp.311-336. Springer-Verlag, Vienna-New York.

\bibitem{belavkin-filtering}
V. P. Belavkin.
\newblock Quantum stochastic calculus and quantum nonlinear
filtering.
\newblock Journal of Multivariate Analysis, 42:171-201,1992.

\bibitem{bouten-sse}
L. Bouten, M. Guta and H. Maassen. \newblock Stochastic
Schrodinger equations. \newblock  MATH.GEN., {\bf 37}, 3189, 2004


\bibitem{Deng1}
H.~Deng and M.~Krstic.
\newblock Stochastic nonlinear stabilization. 1. A backstepping design.
\newblock Systems \& Control Lett. {\bf 32}:151--159, 1997.

\bibitem{Deng2}
H.~Deng and M.~Krstic.
\newblock Stochastic nonlinear stabilization. 2. Inverse optimality.
\newblock Systems \& Control Lett. {\bf 32}:143--150, 1997.

\bibitem{Florchinger2}
P.~Florchinger.
\newblock A stochastic version of {J}urdjevic-{Q}uinn theorem.
\newblock {\em Stochastic Anal. Appl.}, 12:473--480, 1994.

\bibitem{Florchinger1}
P.~Florchinger.
\newblock Lyapunov-like techniques for stochastic stability.
\newblock {\em SIAM J. Control and Optimization}, 33:1151--1169, 1995.

\bibitem{Florchinger3}
P.~Florchinger.
\newblock Feedback stabilization of affine in the control stochastic
  differential systems by the control Lyapunov function method.
\newblock {\em SIAM J. Control and Optimization}, 35:500--511, 1997.


\bibitem{Frankel1}
T.~Frankel.
\newblock {\em The Geometry of Physics: An Introduction}.
\newblock Cambridge University Press, 2nd edition, 1999.

\bibitem{mabuchi-science}
J. M. Geremia and J. K. Stockton and H. Mabuchi.
\newblock Real-time quantum feedback control of atomic
spin-squeezing. \newblock {\em Science}, 304:270-273, 2004.

\bibitem{Guillemin1}
V.~Guillemin and A.~Pollack
\newblock {\em Differential topology}.
\newblock Prentice-Hall Inc., Englewood Cliffs, N.J., 1974.


\bibitem{holevo}
A.~Holevo.
\newblock {\em Statistical Structure of Quantum Theory}.
\newblock Lecture Notes in Physics; Monographs: 67. Springer Verlag, 2001.

\bibitem{Jurdjevic5}
V.~Jurdjevic and J.~P. Quinn.
\newblock Controllability and stability.
\newblock {\em Journal of Differential Equations}, 28:381--389, 1978.

\bibitem{Khasminskiy1}
R. Z. Khas'minskiy.
\newblock {\em Stochastic Stability of Differential Equations}.
\newblock Sijthoff and Noordhoff, Alphen aan den Rijn, The Netherlands, 1980.

\bibitem{Kushner1}
H.~J. Kushner.
\newblock Stochastic stability.
\newblock In R.~Curtain, editor, {\em Stability of stochastic dynamical
  systems}, Lecture Notes in Math., Vol. 294, pages 97--124. Springer, Berlin,
  1972.


\bibitem{maassen-qp}
H. Maassen.
\newblock Quantum probability applied to the damped harmonic
oscillator,
\newblock in {\em Quantum Probability Communications} {\bf XII} 23-58,
eds. S. Attal, J.M. Lindsay, World Scientific, Singapore 2003.


\bibitem{meyer}
P.~A. Meyer.
\newblock {\em Quantum Probability for Probabilists}.
\newblock Lecture Notes in Mathematics 1538. Springer Verlag, 1995.

\bibitem{Mirrahimi2}
M.~Mirrahimi, P.~Rouchon, and G.~Turinici.
\newblock Lyapunov control of bilinear {S}chr{\"o}dinger equations.
\newblock {\em Automatic, in press}, 2005.



\bibitem{Nielsen1}
M.~A. Nielsen and I.~L. Chuang.
\newblock {\em Quantum Computation and Quantum Information}.
\newblock Cambridge Univ. Press, 2000.


\bibitem{parthasarathy}
K.~R. Parthasarathy.
\newblock {\em An Introduction to Quantum Stochastic Calculus}, volume~85 of
  {\em Monographs in Mathematics}.
\newblock Birkhauser, 1992.

\bibitem{sakurai}
J.J. Sakurai.
\newblock {\em Modern Quantum Mechanics}.
\newblock Addison-Wesley, revised edition, 1994.


\bibitem{mabuchi-quantumclassical}
A. C. Doherty, S. Habib, K. Jacobs, H. Mabuchi, S. M. Tan.
\newblock Quantum Feedback Control and Classical Control Theory.
\newblock Phys. Rev. A, {\bf 62}, 012105, 2000 .

\bibitem{vanhandel} R. van Handel, J. K.
Stockton and H. Mabuchi. \newblock Feedback Control of Quantum
State Reduction. \newblock IEEE Trans. Automat. Control, {\bf 50},
768-780, 2005.

\bibitem{wang1}
J.~Wang and H.~M.~Wiseman.
\newblock Feedback-stabilization of an arbitrary pure state of a two-level atom.
\newblock Phys. Rev. A {\bf 64}, 063810, 2001.

\bibitem{wiseman-bayesian}
  H. M. Wiseman, S. Mancini and J. Wang.
  \newblock Bayesian feedback versus Markovian feedback in a two-level atom,
  \newblock Phys. Rev. A {\bf 66}, 013807, 2002.

\bibitem{wiseman-milburn} H. M. Wiseman and G. J. Milburn.
\newblock Quantum theory of optical feedback via homodyne
detection.
\newblock Phys. Rev. Lett.,{\bf 70}:5, 548-551, 1993.

\bibitem{Zyczkowski2}
K. Zyczkowski and W. Slomczy{\'n}ski.
\newblock Monge metric on the sphere and geometry of quantum states.
\newblock J. Phys. A, {\bf 34}:6689, 2000.



\end{thebibliography}

\end{document}